Wenshan Chen ✉ ; Kingsley Egbo ; Joe Kler ; Andreas Falkenstein ; Jonas Lähnemann ;
Oliver Bierwagen

View Online / Export Citation

---

### Articles You May Be Interested In

Structural and vibrational properties of Sn$_x$Ge$_{1-x}$: Modeling and experiments

*J. Appl. Phys.* (July 2018)

*In situ* study and modeling of the reaction kinetics during molecular beam epitaxy of GeO$_2$ and its etching by Ge

*APL Mater.* (July 2023)

Etching of elemental layers in oxide molecular beam epitaxy by O$_2$-assisted formation and evaporation of their volatile (sub)oxide: The examples of Ga and Ge

*J. Vac. Sci. Technol. A* (April 2024)

# Kinetics, thermodynamics, and catalysis of the cation incorporation into GeO$_2$, SnO$_2$, and (Sn$_x$Ge$_{1-x}$)O$_2$ during suboxide molecular beam epitaxy

Wenshan Chen,[1,a)] Kingsley Egbo,[1,b)] Joe Kler,[2] Andreas Falkenstein,[2] Jonas Lähnemann,[1] and Oliver Bierwagen[1,c)]

## AFFILIATIONS

[1] Paul-Drude-Institut für Festkörperelektronik, Leibniz-Institut im Forschungsverbund Berlin e.V., Hausvogteiplatz 5–7, D-10117 Berlin, Germany
[2] Institute of Physical Chemistry, RWTH Aachen University, Aachen, Germany

Note: This paper is part of the Special Topic on Ultrawide Bandgap Semiconductors.
a) Author to whom correspondence should be addressed: chen@pdi-berlin.de
b) Present Address: National Renewable Energy Laboratory, Golden, Colorado 80401, USA
c) Electronic mail: bierwagen@pdi-berlin.de

## ABSTRACT

Rutile GeO$_2$ is a promising ultra-wide bandgap semiconductor for future power electronic devices whose alloy with the wide bandgap semiconductor rutile-SnO$_2$ enables bandgap engineering and the formation of heterostructure devices. The (Sn$_x$Ge$_{1-x}$)O$_2$ alloy system is in its infancy, and molecular beam epitaxy (MBE) is a well-suited technique for its thin-film growth, yet it presents challenges in controlling the alloy composition and growth rate. To understand and mitigate this challenge, the present study comprehensively investigates the kinetics and thermodynamics of suboxide incorporation into GeO$_2$, SnO$_2$, and (Sn$_x$Ge$_{1-x}$)O$_2$ during suboxide MBE (S-MBE), the latest development in oxide MBE using suboxide sources. We find S-MBE to simplify the growth kinetics, offering better control over growth rates than conventional MBE but without supporting cation-driven oxide layer etching. During binary growth, SnO incorporation is kinetically favored due to its higher oxidation efficiency and lower vapor pressure (limiting its loss by desorption) compared to those of GeO. In (Sn$_x$Ge$_{1-x}$)O$_2$ growth, however, the GeO incorporation is preferred and the SnO incorporation is suppressed, indicating a catalytic effect, where SnO promotes GeO incorporation. The origin of this catalytic effect cannot be understood by comparing the binary kinetics or thermodynamics (cation–oxygen bond strengths), thus calling for further theoretical studies. Our experimental study provides guidance for controlling the growth rate and alloy composition of (Sn$_x$Ge$_{1-x}$)O$_2$ in S-MBE, highlighting the impact of the substrate temperature and active oxygen flux besides that of the mere SnO:GeO flux stoichiometry. The results are likely transferable to further physical and chemical vapor deposition methods, such as conventional and hybrid MBE, pulsed laser deposition, mist-, or metalorganic chemical vapor deposition.

## I. INTRODUCTION

Ultrawide bandgap (UWBG) semiconductors are gaining significant attention for their potential in next-generation power electronics.[1–5] Among these materials, rutile GeO$_2$ (r-GeO$_2$, E$_g$ ≈ 4.7 eV) stands out due to its superior predicted Baliga's figure of merit compared to $\beta$-Ga$_2$O$_3$, SiC, and GaN,[6] as well as larger thermal conductivity compared to $\beta$-Ga$_2$O$_3$.[7] In addition, r-GeO$_2$ has been predicted to be bipolar dopable,[8] a rare and valuable property among oxide materials, which further enhances its attractiveness for electronic applications. Despite a predicted electron mobility of >300cm$^2$/Vs,[6] reports of the measured electron transport properties of r-GeO$_2$ are so far indicating low room temperature mobilities (<10 cm$^2$/Vs) that do not allow it to be extracted by Hall measurements[9]—possibly related to a low material quality. The rutile form of SnO$_2$ (E$_g$ ≈ 3.6 eV), closely related to r-GeO$_2$, has been extensively studied and shows controllable electron transport properties with room temperature mobilities exceeding 200 cm$^2$/Vs.[10–14] When alloyed with GeO$_2$, the formed (Sn$_x$Ge$_{1-x}$)O$_2$ potentially combines the aforementioned superior characteristics of GeO$_2$ with the outstanding electronic properties traditionally associated with SnO$_2$. For instance, an unintentionally doped r-Ge$_{0.49}$Sn$_{0.51}$O$_2$ film with a carrier density of $7.8 \times 10^{18}$ cm$^{-3}$ and a mobility of 24 cm$^2$ V$^{-1}$ s$^{-1}$ has been reported.[15] In addition, 1 at. % Ta-doped r-Ge$_x$Sn$_{1-x}$O$_2$ thin films with Ge content $x \leq 0.36$ demonstrated a low electrical resistivity of $2-3 \times 10^{-4}$ Ω cm, attributed to their high carrier density ($>3 \times 10^{20}$ cm$^{-3}$) and effective electron Hall mobility ($\approx 70$ cm$^2$V$^{-1}$s$^{-1}$).[16] This combination makes (Sn$_x$Ge$_{1-x}$)O$_2$ an attractive material system for advanced electronic applications, particularly in bandgap engineering and the development of heterostructure devices.

High purity, crystallinity, and compositional control are essential for these applications and can be achieved through molecular beam epitaxy (MBE). MBE offers a relatively simple mechanism for growing high-quality oxide layers. However, despite the successful growth of phase-pure, high-quality r-SnO$_2$,[11,14,17,18] MBE is challenged by the limited growth rates for r-GeO$_2$ related to the desorption of the intermediately formed high-vapor-pressure suboxide GeO.[19–21]

GeO$_2$ is furthermore prone to forming amorphous and α-quartz phase due to its intrinsic material properties.[22] The strong covalent bonding and flexible network structure favor the formation of the amorphous (glass) phase, while the thermodynamically stable tetrahedral coordination is characteristic of the α-quartz phase. These factors complicate the synthesis of the desired rutile phase of GeO$_2$.[23] Despite successful r-GeO$_2$ synthesis using various thin film growth methods, achieving higher growth rates and phase purity remains a current challenge.[19,24,25]

The miscibility of GeO$_2$ and SnO$_2$ for the ternary (Sn$_x$Ge$_{1-x}$)O$_2$ system has been reported,[26] being a potential avenue for stabilization of the rutile phase. The growth of (Sn$_x$Ge$_{1-x}$)O$_2$ shows significant potential, with successful epitaxial r-(Sn$_{0.47}$Ge$_{0.53}$)O$_2$ thin films achieved by using mist chemical vapor deposition (mist-CVD).[15] In addition, Ta-doped r-(Sn$_x$Ge$_{1-x}$)O$_2$ thin films with Ge content $0 \leq (1-x) \leq 0.7$ have been grown by pulsed laser deposition (PLD).[16] The complex control of (Sn$_x$Ge$_{1-x}$)O$_2$ alloy compositions by MBE has so far only been demonstrated in hybrid MBE (using an organic Sn-precursor) for Ge-contents $1 - x \leq 0.54$.[27]

To date, there is a lack of systematic investigations into the reaction kinetics and thermodynamics of the Ge and Sn incorporation for the growth of (Sn$_x$Ge$_{1-x}$)O$_2$. Understanding these aspects is crucial for optimizing the growth process and achieving high-quality thin films with precise control over alloy composition and growth rate.

In this study, we systematically investigate the kinetics (rates) and thermodynamics (theoretical change in Gibbs free energy for the associated chemical reactions) of cation incorporation in GeO$_2$, SnO$_2$, and (Sn$_x$Ge$_{1-x}$)O$_2$, although without considering the impact of different structural phases. First, we compare the growth kinetics of conventional MBE to that of suboxide MBE (S-MBE) for GeO$_2$ and SnO$_2$. Next, we compare the quantitative differences of growth kinetics between binary GeO$_2$ and SnO$_2$, focusing particularly on oxidation efficiency (the fraction of provided oxygen that participates in the oxidation reaction) and the growth rate dependence on temperature. Finally, we explore the competing SnO and GeO incorporation into ternary (Sn$_x$Ge$_{1-x}$)O$_2$ during S-MBE as a function of $T_G$ at varying O$_2$ flow rates, using equivalent SnO and GeO suboxide fluxes.

Our results reveal that the conventional MBE and S-MBE offer complementary advantages. Conventional MBE is effective at etching oxide layers under oxygen-free conditions, a capability that S-MBE lacks. Conversely, S-MBE simplifies the reaction kinetics of binary GeO$_2$ and SnO$_2$ from a complex two-step oxidation mechanism observed in conventional MBE to a straightforward single-step process. In the growth of binary GeO$_2$ and SnO$_2$, the incorporation of Sn or SnO into SnO$_2$ is kinetically favored—meaning that the oxidation conditions promote a higher growth rate for SnO$_2$ compared to GeO$_2$ under similar conditions—enabling SnO$_2$ to grow at higher temperatures and lower oxygen flow rates. For the growth of ternary (Sn$_x$Ge$_{1-x}$)O$_2$, in contrast, we observe an enhanced incorporation of GeO in the presence of SnO, coinciding with a suppression of SnO incorporation in the presence of GeO. This behavior, subject to future theoretical investigation, may be driven by thermodynamic interactions between SnO and GeO species or by a kinetic mechanism beyond those observed in binary growth systems.

The miscibility of SnO$_2$ and GeO$_2$, as well as the crystallinity of (Sn$_x$Ge$_{1-x}$)O$_2$, are beyond the scope of this study. Although the studied Ge-containing films were found to be amorphous while the pure SnO$_2$ films were rutile, we anticipate that the kinetic and thermodynamic insights gained here will have broader applicability to both amorphous and crystalline (rutile) phase, as the enthalpy and entropy differences between these phases are relatively small. Thus, we believe this work will inform future studies on the rutile-phase (Sn$_x$Ge$_{1-x}$)O$_2$ thin film growth.

## II. EXPERIMENTAL DETAILS

For this study, binary GeO$_2$, SnO$_2$, and ternary (Sn$_x$Ge$_{1-x}$)O$_2$ layers were grown on 2-in. diameter, single-side polished, c-plane sapphire (Al$_2$O$_3$(0001)) wafers by plasma-assisted MBE. The rough backside of the substrate was sputter-coated with titanium to allow for radiative heating by the substrate heater during growth. The growth temperature ($T_G$) was measured with a thermocouple placed behind the substrate heater.

For the S-MBE, standard shuttered effusion cells with Al$_2$O$_3$ crucibles were used. The source material for the suboxide GeO flux was GeO$_2$ (5N purity) powder following the approach of Ref. 19, whereas that for the SnO flux was a mixture of Sn metal (5N purity) and SnO$_2$ (4N purity), providing a particularly efficient suboxide source.[28] For the conventional MBE, a standard shuttered effusion cell was used to evaporate Ge (7N purity) from a pyrolytic BN crucible. The growth of SnO$_2$ from a Sn source is described in Ref. 29. The beam equivalent pressure (BEP), proportional to the particle flux, was measured by a nude filament ion gauge positioned at the substrate location. The BEPs are given in units of mbar and were converted into the equivalent growth rates of GeO$_2$ and SnO$_2$ (in nm/min). The conversion factor was established by measuring the growth rate (Γ) of the GeO$_2$ and SnO$_2$ layers under conditions of full Ge and Sn incorporation and the related BEP. Full incorporation is ensured by growing in an oxygen-rich regime at a sufficiently low substrate temperature of 300 °C, where desorption of GeO and SnO is negligible. A GeO cell temperature of 1010 °C was maintained throughout the experiments. The relation of $\Phi_{GeO}$ between the measured BEPs is $\Phi_{GeO}$ = 5.3 nm/min ≡ 1.2 × 10$^{-7}$ mbar . For Ge, a cell temperature of 1300 °C resulting in $\Phi_{Ge}$ = 5.3 nm/min ≡ 1.24 × 10$^{-7}$ mbar was used. The corresponding BEPs of GeO and Ge as a function of their cell temperatures are shown in Fig. S1 in the supplementary material. For SnO, due to the variations in the position of Sn metal particles and SnO$_2$ powder in the crucible, the BEPs and cell temperatures do not follow an ideal Arrhenius behavior. Therefore, we set and ensure the desired flux by measuring BEPs and adjusting the SnO cell temperatures before each growth run based on a calibrated relationship of $\Phi_{SnO}$ = 5.3 nm/min ≡ 1.9 × 10$^{-7}$ mbar. A radio frequency plasma source with a mass flow controller supplied an activated oxygen flux from the research-grade O$_2$ gas (6N purity). The radio frequency power of the plasma source was maintained at 300 W in all experiments in this study. The O$_2$ mass flow is given in standard cubic centimeter per minute (SCCM), where 1 SCCM corresponds to an O$_2$ BEP and flux at the substrate of ≈10$^{-5}$ mbar and ≈3 × 10$^{15}$ cm$^{-2}$s$^{-1}$, respectively.[30] In situ laser reflectometry (LR) with a laser emitting at a wavelength of 405 nm at an incident angle of 30° (60° with respect to the substrate normal) was used to extract the growth and etch rates for all layers.

X-ray diffraction and reflection high-energy electron diffraction indicated the formation of rutile-SnO$_2$(100) and amorphous GeO$_2$ and (Sn$_x$Ge$_{1-x}$)O$_2$ layers, prompting us to assume corresponding refractive indices and cation number densities of 1.6 and 2.1 × 10$^{22}$ cm$^{-3}$ for GeO$_2$,[20,31] 2.1 and 2.8 × 10$^{22}$ cm$^{-3}$ for SnO$_2$,[32,33] and—as their average refractive index—1.8 for (Sn$_x$Ge$_{1-x}$)O$_2$.

The Sn (x) and Ge (1 − x) concentrations in the (Sn$_x$Ge$_{1-x}$)O$_2$ layers were determined by time-of-flight secondary ion mass spectrometry (ToF-SIMS) calibrated by an Sn$_{0.45}$Ge$_{0.55}$O$_2$ sample whose x was determined by scanning electron microscope energy dispersive x-ray spectroscopy (EDX). The ToF-SIMS measurements were performed using a ToF-SIMS IV machine (IONTOF GmbH, Münster, Germany), equipped with a ToF-SIMS V analyzer and an extended dynamic range (EDR) unit. A low energy (O$_2$)$^+$ was used for sputter etching the sample and a high-energy Ga$^+$ gun for producing secondary ions for ToF analysis. The Ga$^+$ gun was operated with an ion energy of 25 keV and an analysis raster of 100 μm × 100 μm. The (O$_2$)$^+$ gun was operated at 2 keV and rastered over an area of 300 μm × 300 μm. Charge compensation was achieved by flooding the surface with low-energy electrons (<20 eV). Positive secondary ions were detected with a cycle time of 50 μs. ToF-SIMS craters were measured post-analysis by interference microscopy (NT1100, Veeco Instruments Inc., NY, USA).

## III. RESULTS AND DISCUSSIONS
### A. *In situ* oxide etching

Prior MBE experiments have demonstrated that oxide layers, such as In$_2$O$_3$, Ga$_2$O$_3$, SnO$_2$, and GeO$_2$, can be etched in the absence of oxygen by a flux of their respective cations (In, Ga, Sn, and Ge),[20,34] and In$_2$O$_3$ can additionally be etched by a Ga flux.[35] The underlying mechanism is the formation and evaporation of volatile suboxides as schematically shown in Fig. 1(a). To extend this framework, we investigate the capability of Ge and SnO fluxes to etch an SnO$_2$ layer *in situ*.

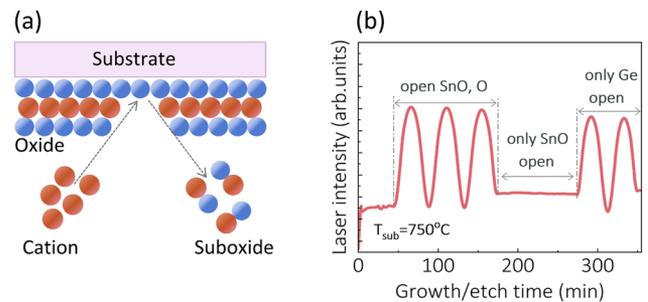

FIG. 1. *In situ* oxide etching in MBE. (a) Schematic describing the oxide etching process by providing cation flux in an oxygen-free environment. Red symbols represent the cations and blue symbols represent the oxygen atoms. (b) Laser reflectometry signal during the SnO$_2$ layer growth and its subsequent etching attempts by SnO and Ge flux. The events of shutter opening and closing are indicated in the figure.

Our experiment involved initially growing a SnO$_2$ layer at a substrate temperature ($T_G$) of 750 °C with an O$_2$ flow rate of 0.5 SCCM. Following deposition, we transitioned to an etching phase in ultra-high vacuum without provision of an oxygen flux. The SnO$_2$ layer was exposed first to the SnO flux ($\Phi_{SnO}$ = 3 nm/min) and, subsequently, to a Ge flux ($\Phi_{Ge}$ = 5.3 nm/min) while maintaining the $T_G$ constant at 750 °C. Figure 1(b) presents the reflected laser intensity measured by laser reflectometry (LR) as a function of time during the SnO$_2$ growth and its subsequent exposure to pure SnO and Ge fluxes. The LR signal remained unchanged during the exposure of SnO$_2$ to SnO flux alone, indicating that SnO does not etch the SnO$_2$ layer. In contrast, a significant change in the LR signal was observed once we switched from opening the SnO shutter to opening the Ge shutter. The observed discontinuous transition in the oscillations signifies a reduction in the film thickness, implying that Ge effectively etched the SnO$_2$ through the reaction

$$\text{Ge } (a) + \text{SnO}_2 \ (s) \rightarrow \text{GeO } (g) + \text{SnO } (g), \quad (1)$$

where *a*, *g*, and *s* denote the adsorbate, gaseous, and solid phases, respectively. The observed oscillatory periods during growth and etching allow us to determine a growth rate of 3 nm/min and an etch rate of 5.3 nm/min. This etch rate aligns with the prediction in Eq. (1), indicating that the etching reaction took place. In addition, we calculated the temperature-dependent Gibbs free energy changes for Eq. (1) in the supplementary material. A resulting negative Gibbs free energy change indicates that the reaction is thermodynamically feasible.

These etching reactions highlight the advantages of conventional MBE, particularly regarding the reductive etching of semiconducting oxide materials not only by their constituent cations but also by particular foreign cations. The suboxide, in contrast, is not able to etch its related oxide. The ability to exploit reactions between cations and oxides offers opportunities for selective engineering of oxide etching processes without the necessity for additional reactive gases.

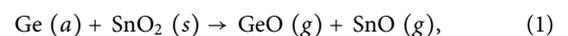

### B. Growth mechanism of conventional vs suboxide MBE

When using an elemental cation flux to grow oxide materials (e.g., $In_2O_3$, $Ga_2O_3$, $SnO_2$, and $GeO_2$), the corresponding conventional MBE growth kinetics is governed by a two-step oxidation mechanism.[20,36] The mechanism involves the intermediate formation of suboxides, whose desorption naturally limits the rate of oxide growth and leads to non-intuitive flux dependences of the growth rate. In S-MBE, in contrast, a suboxide flux is directly provided from the source, thus omitting the suboxide formation step on the substrate.

These scenarios are schematically illustrated in Figs. 2(c) and 2(d) along with the related Γ of $GeO_2$ as a function of $O_2$ flow rate at $T_G$ = 600 °C from both S-MBE and conventional MBE under equivalent GeO and Ge fluxes in Fig. 2(a). Without oxygen flux, the Γ is negative due to layer etching in conventional MBE[20] [schematically shown in Fig. 1(a)] but zero in S-MBE as the suboxide GeO cannot decompose the $GeO_2$. In S-MBE, Γ proportionally increases with the $O_2$ flow rate in the suboxide-rich regime until it reaches the stoichiometric point, corresponding to an $O_2$ flow rate of 0.65 SCCM for the present GeO flux, where all available GeO and activated O atoms are fully incorporated into the film. Beyond this stoichiometric point, the Γ plateaus in the O-rich regime despite further increases in the $O_2$ flow rate as it is limited by the GeO-flux.

In contrast, when using conventional MBE with a Ge flux equivalent to the one chosen above for S-MBE, the onset of film growth, transitioning from etching into the Ge-rich regime, happens at a finite $O_2$ flow rate of 0.5 SCCM, and the full oxidation of the Ge-flux into $GeO_2$ requires a higher $O_2$ flow rate of 1.2 SCCM. This behavior is explained by the two-step oxidation kinetics mentioned above for MBE-grown oxides that possess suboxides with higher vapor pressure than their cation element [Fig. 2(c)].[20,36] In the first step, Ge is oxidized to the suboxide GeO (Ge + O → GeO), followed by the second step, in which GeO is further oxidized to $GeO_2$ (GeO + O → $GeO_2$)—provided there is a sufficient supply of active O. A portion of the GeO generated in the first oxidation step desorbs if there is an insufficient O flux for the second oxidation step or $T_G$ is too high. The extrapolated $O_2$ flow rate of 0.5 SCCM marks the threshold to $GeO_2$ growth, indicating the stoichiometric formation of GeO, which corresponds to the growth onset at 0 SCCM in S-MBE growth. To achieve stoichiometric $GeO_2$ growth, an additional $O_2$ flow of ∼0.7 SCCM, yielding a total of 1.2 SCCM, is necessary to oxidize all formed GeO. This aligns with the $O_2$ flow rate of ≈0.65 SCCM required in S-MBE to ensure the complete oxidation of the provided GeO flux to $GeO_2$.

We observed the same qualitative differences between conventional and S-MBE for $SnO_2$ growth as shown in Fig. 2(b). In the following, we will discuss the quantitative differences between $GeO_2$ and $SnO_2$ growth in detail.

### C. Quantitative growth kinetics comparison of binary $GeO_2$ vs $SnO_2$ by conventional and suboxide MBE

Quantitative comparison of the growth kinetics of $GeO_2$ and $SnO_2$ by conventional MBE with equivalent Ge and Sn fluxes (Fig. 2(a) vs (b)) indicates that less active oxygen is required to form SnO than GeO (see the transition from etching to cation-rich growth regime at 0.15 vs 0.5 SCCM) and to form $SnO_2$ than $GeO_2$ (transition to O-rich growth plateau at 0.36 vs 1.2 SCCM). Next, we quantitatively compare the growth kinetics of $GeO_2$ and $SnO_2$ by S-MBE with equivalent GeO and SnO fluxes. Figure 3(a) summarizes the $O_2$ flow dependent Γ of $GeO_2$ and $SnO_2$ from Fig. 2. At $T_G$ = 700 °C, the stoichiometric growth for $SnO_2$ happens at an $O_2$ flow rate of 0.17 SCCM, while for $GeO_2$ it increases to 0.84 SCCM due to oxygen-deficiency-induced desorption of the volatile suboxide GeO. Notably, the $O_2$ flow rate required for stoichiometric growth of $GeO_2$ decreases to 0.65 SCCM when the $T_G$ is reduced to 600 °C, suggesting an enhanced oxidation efficiency of GeO at lower $T_G$, which is still well below that of SnO.

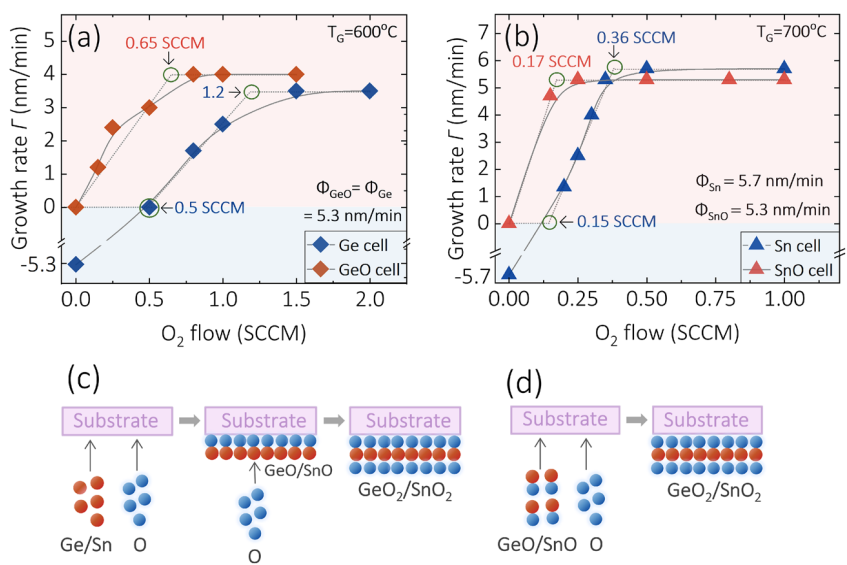

**FIG. 2.** Measured growth rates (Γ) of (a) $GeO_2$ and (b) $SnO_2$ as a function of $O_2$ flow rate in S-MBE (using suboxide flux) and conventional MBE (using elemental cation flux). The applied suboxide and elemental cation fluxes, along with the corresponding growth temperatures ($T_G$), are indicated in the figure. Open symbols denote the stoichiometric growth conditions for each method, with the corresponding $O_2$ flow rates specified. The data for Sn flux are from Ref. 29. (c) and (d) Schematics describing the growth scenarios: (c) conventional MBE with Ge/Sn atomic flux, where the binary oxide form through a two-step oxidation and (d) S-MBE with GeO/SnO suboxide flux, where the binary form in a single oxidation step.

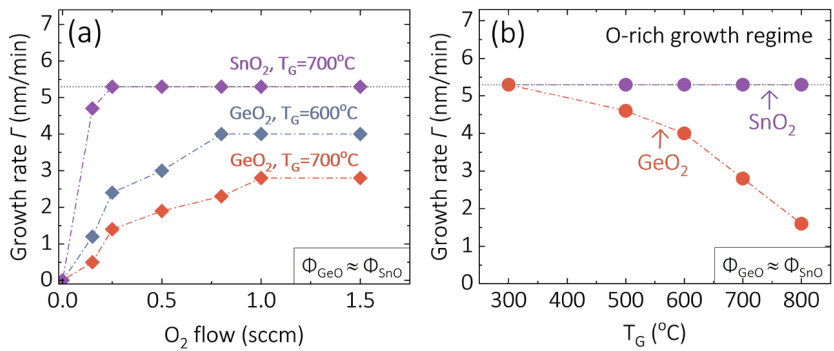

**FIG. 3.** (a) $\Gamma$ of $GeO_2$ and $SnO_2$ as a function of $O_2$ flow rate using S-MBE. The corresponding $T_G$ are indicated in the figure. (b) $T_G$ dependent $\Gamma$ of $GeO_2$ and $SnO_2$ in S-MBE in O-rich regime (1 SCCM for $GeO_2$ and 0.5 SCCM for $SnO_2$) with equivalent GeO and SnO flux of 5.3 nm/min as denoted by the dotted line.

**TABLE I.** Total (molecular and atomic) oxygen flux, oxidized cation ("cat.") and suboxide ("subox.") fluxes at stoichiometric points indicated in Figs. 2(a) and 2(b) for suboxide or oxide formation, and corresponding oxidation efficiency (fraction of total provided oxygen contributing to the oxidation). "GR" and "$N_{cat.}$" denote the growth rate and cation number density of the films; their product is the "flux" of cation or suboxide oxidized and, thus, equals the flux of utilized O atoms for this oxidation.

| Oxidation step | $O_2$ flow (SCCM) | Total $O_x$ flux (O nm$^{-2}$ s$^{-1}$) | GR (nm/min) | $N_{cat.}$ (nm$^{-3}$) | Oxidized cat./subox. (nm$^{-2}$ s$^{-1}$) | Oxidation efficiency (%) |
|---|---|---|---|---|---|---|
| MBE: Ge + O → GeO | 0.5 | 30 | 5.3 | 21 | 1.9 | 6.3 |
| MBE: GeO + O → $GeO_2$ | 0.7 | 41 | 3.5 | 21 | 1.2 | 3.0 |
| S-MBE: GeO + O → $GeO_2$ | 0.65 | 38 | 4.0 | 21 | 1.4 | 3.7 |
| MBE: Sn + O → SnO | 0.15 | 8.9 | 5.7 | 28 | 2.7 | 30 |
| MBE: SnO + O → $SnO_2$ | 0.21 | 12 | 5.7 | 28 | 2.7 | 21 |
| S-MBE: SnO + O → $SnO_2$ | 0.17 | 10 | 5.3 | 28 | 2.5 | 25 |

Table I uses these results to estimate the "oxidation efficiencies" as a ratio of oxygen contributing to the different oxidation steps for the Ge- and Sn-systems (at 600 and 700 °C, respectively) over the total provided oxygen flux. For conventional and S-MBE, a 1.5–2 times higher oxidation efficiency for the cation oxidation compared to suboxide oxidation can be found, and a 5–6 times higher oxidation efficiency (20%–30%) for the Sn-system compared to the Ge-system (3%–6%).

Figure 3(b) further compares the measured $\Gamma$ of $GeO_2$ and $SnO_2$ grown in the O-rich regime by S-MBE as a function of $T_G$ ranging from 300 to 800 °C. For $SnO_2$, $\Gamma$ remains constant across the entire range of $T_G$. In contrast, for $GeO_2$, there is a strong decrease in $\Gamma$ with increasing $T_G$ due to a pronounced desorption of the GeO suboxide. This difference can be explained kinetically by the higher vapor pressure of GeO compared to that of SnO (shown in Fig. S2 in the supplementary material).

Thus, higher oxidation efficiency and lower vapor pressure of SnO compared to GeO indicate a kinetically preferred formation of binary $SnO_2$ compared to $GeO_2$.

### D. SnO-catalyzed incorporation of GeO in $(Sn_xGe_{1-x})O_2$ growth

To investigate the competing cation incorporation during ternary growth, amorphous $(Sn_xGe_{1-x})O_2$ multilayers were grown with nominally equal GeO and SnO fluxes ($\Phi_{GeO} = \Phi_{SnO}$ = 5.3 nm/min)[37] while $T_G$ and $O_2$ flow rates were varied from layer to layer. With this approach we are able to investigate the impact of growth conditions on cation incorporation by the growth and characterization of only one multi-layer sample. Figure S3 of the supplementary material shows the LR signal during growth, which allows us to determine ternary $\Gamma_{(Sn_xGe_{1-x})O_2}$.

Figure 4(a) presents the $\Gamma$ of the ternary $(Sn_xGe_{1-x})O_2$ thin film as a function of $T_G$ at $O_2$ flow rates of 0.1 and 0.25 SCCM. Lower $T_G$ and higher $O_2$ flow rates facilitate faster growth as they suppress parasitic GeO and SnO desorption, a behavior already known from the growth of the binary oxides. The corresponding Ge composition $(1 - x)$, determined by time-of-flight secondary ion mass spectrometry (ToF-SIMS) is shown in Fig. 4(b). (Detailed ToF-SIMS results and calibrations are provided in Fig. S4 of the supplementary material.) In addition, EDX measurements confirm the sample's oxygen stoichiometry, as shown in Fig. S5 in the supplementary material. Surprisingly, the Ge composition of the layer is well above that of the fluxes (dotted horizontal line). In stark contrast to the kinetically preferred oxidation of SnO in binary growth, this result suggests a preferred oxidation of GeO during ternary growth. While the Ge composition is highly sensitive to $T_G$ at 0.25 SCCM, it shows limited sensitivity at 0.1 SCCM.

For a quantitative discussion of the individual incorporation rates of SnO and GeO into $(Sn_xGe_{1-x})O_2$, we extract the equivalent pseudo-binary growth rates $\gamma$ of $SnO_2$ and $GeO_2$ from the cation compositions and total ternary growth rate shown in Fig. 4 as $\gamma_{SnO_2} = x \, \Gamma_{(Sn_xGe_{1-x})O_2}$, and $\gamma_{GeO_2} = (1 - x) \, \Gamma_{(Sn_xGe_{1-x})O_2}$, respectively.[38] The results are shown in Fig. 5(a) in comparison to the binary incorporation rate $\Gamma$ as a function of the $T_G$ at an $O_2$ flow rate of 0.1 SCCM. Although in binary growth the SnO incorporation is kinetically favored over that of GeO, our observations reveal a surprisingly enhanced incorporation of GeO ($\gamma_{GeO_2} \approx 6.6 \times \Gamma_{GeO_2}$) in the presence of SnO. Conversely, the presence of GeO significantly suppresses the incorporation of SnO ($\gamma_{SnO_2} \approx 0.15 \times \Gamma_{SnO_2}$).

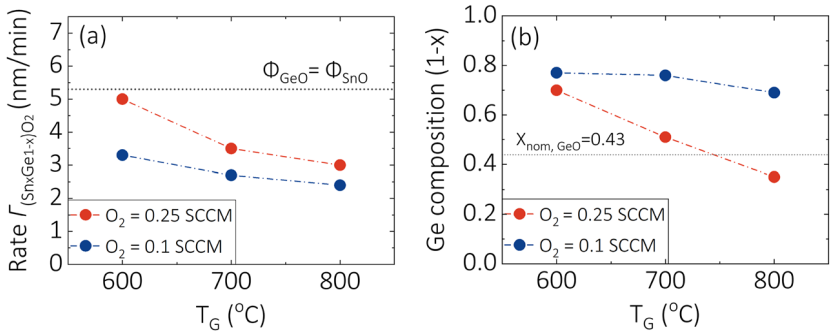

**FIG. 4.** (a) Alloy growth rate $\Gamma_{(Sn_xGe_{1-x})O_2}$ as a function of $T_G$ with $\Phi_{GeO} = \Phi_{SnO}$ for varied $O_2$ flow of 0.25 SCCM and 0.1 SCCM. (b) The corresponding Ge composition $(1 - x)$ of the $(Sn_xGe_{1-x})O_2$ samples shown in Fig. 3(a). Dotted reference lines indicate the fluxes as equivalent binary growth rates (a) and the flux stoichiometry $1 - x = GeO/(SnO + GeO)$ (b).

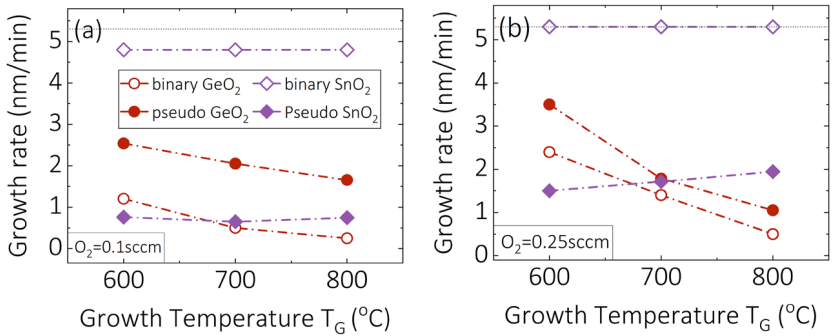

**FIG. 5.** The comparison of binary growth rate $\Gamma$ and pseudo binary growth rate $\gamma$ for $GeO_2$ and $SnO_2$ as a function of growth temperature $T_G$ for $\Phi_{GeO} = \Phi_{SnO}$ at $O_2$ flow rates of (a) 0.1 SCCM and (b) 0.25 SCCM.

These contrasting results between the binary incorporation rate $\Gamma$ and the pseudo-binary incorporation rate $\gamma$ lead us to hypothesize that the incorporation of GeO into $(Sn_xGe_{1-x})O_2$ is thermodynamically more favorable than that of SnO in analogy to the case of $(In_xGa_{1-x})_2O_3$ growth[35] and is catalyzed by cation exchange.[39] This behavior is particularly pronounced at the low $O_2$ flow rate of 0.1 SCCM, where the environment is insufficient for the complete oxidation of either GeO or SnO.

Here, we follow the proposed mechanism of suboxide-catalyzed MBE growth applied to $Ga_2O_3$ and $In_2O_3$ by Vogt et al.,[40] schematically shown for our case in Figs. 6(a)–6(c). In a first step, the SnO adsorbate undergoes a kinetically favored oxidation, reacting with the O-flux to form a SnO–O adsorbate that adheres to the surface,

$$SnO(a) + O(a) \rightarrow SnO-O(a). \quad (2)$$

The interaction between SnO and O is through physisorption, characterized by weak van der Waals forces. The presence of GeO destabilizes the SnO–O. This encourages the dissociation of the SnO–O adsorbate and the subsequent oxidation of GeO, which can be described by the "catalytic GeO oxidation" reaction,

$$SnO-O(a) + GeO(a) \rightarrow GeO_2(s) + SnO(a). \quad (3)$$

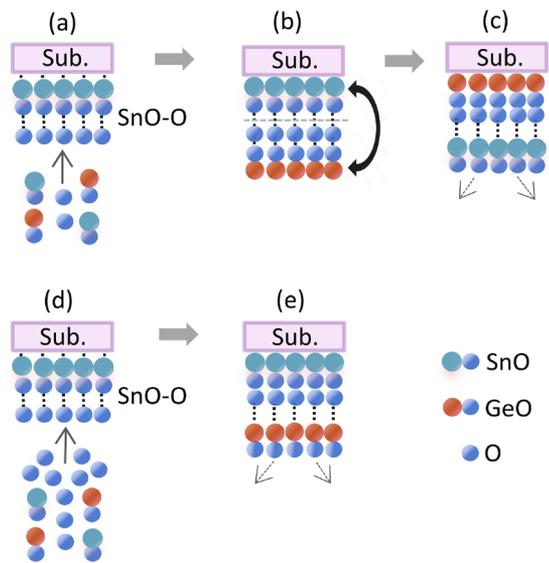

**FIG. 6.** Schematics illustrating the limiting growth mechanisms of ternary $(Sn_xGe_{1-x})O_2$ films by S-MBE under different oxygen-fluxes. (a)–(c) SnO-catalyzed GeO oxidation under low-oxygen conditions: (a) SnO–O adsorbates form on the substrate surface, (b) GeO replaces SnO on the surface, and (c) GeO oxidizes to form $GeO_2$, releasing adsorbed SnO. (d)–(e) Catalyst passivation under oxygen-rich conditions: (d) SnO–O adsorbates form on the substrate surface; and (e) SnO is oxidized to $SnO_2$, adsorbed GeO desorbs.

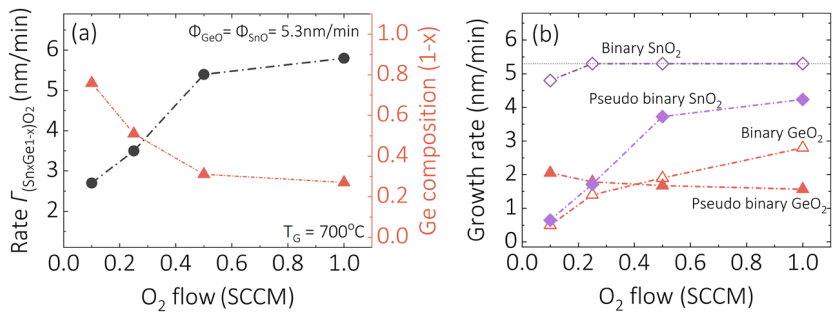

**FIG. 7.** (a) Ternary $\Gamma_{(Sn_xGe_{1-x})O_2}$ and the corresponding Ge composition as a function of $O_2$ flow rate at $T_G = 700\,°C$ for $\Phi_{GeO} = \Phi_{SnO} = 5.3$ nm/mim. (b) The comparison of binary $\Gamma$ from Fig. 3(a) and pseudo-binary $\gamma$ extracted from Fig. 7(a) for $GeO_2$ and $SnO_2$; the dotted line refers to $\Phi_{GeO}$ and $\Phi_{SnO}$.

The SnO(a) produced from this reaction desorbs or reenters the catalytic cycle and can be re-oxidized to SnO–O in reaction (2), which in turn facilitates the further oxidation of GeO in reaction (3). Thus, the low $O_2$ flow promotes a catalytic mechanism for SnO that enhances the incorporation of GeO while concurrently suppressing the incorporation of SnO in the resulting ternary $(Sn_xGe_{1-x})O_2$.

We observed a decrease in Ge composition at higher $O_2$ flow rates, as illustrated in Fig. 4(b). In particular, Fig. 5(b) shows that the enhancement of GeO incorporation diminishes to ($\gamma_{GeO_2} \approx 2 \times \Gamma_{GeO_2}$) while the incorporation of Sn increases to ($\gamma_{SnO_2} \approx 0.33 \times \Gamma_{SnO_2}$) as the $O_2$ flow rate increases from 0.1 to 0.25 SCCM. To further understand these dynamics, Fig. 7(a) shows the $\Gamma$ of ternary $(Sn_xGe_{1-x})O_2$ and the corresponding Ge composition $(1-x)$ across a broader range of $O_2$ flow rates at $T_G = 700\,°C$, and Fig. 7(b) provides a comparative analysis of the $O_2$ flow rate dependent binary $\Gamma$ and the pseudo-binary $\gamma$ for $GeO_2$ and $SnO_2$. This figure indicates that the catalytic effect of SnO on GeO incorporation diminishes further with increasing $O_2$ flow rate. At an $O_2$ flow rate of $\approx 0.4$ SCCM, the enhancement of GeO incorporation ceases entirely, with SnO incorporation being favored over GeO. This behavior at elevated $O_2$ flow can be attributed to the passivation of the catalyst SnO–O layer by the following "catalyst passivation" reaction before dissociation by GeO occurs:

$$SnO(g) + O(g) \rightarrow SnO_2(s) \quad (4)$$

and incorporation of the formed $SnO_2$ into the film. A brief schematic describing this process is shown in Figs. 6(d) and 6(e).

### 1. Thermodynamics considerations

To quantify the underlying thermodynamics, thermochemical calculations (see the supplementary material for the related calculation details) were performed to compare the Gibbs free energy change ($\Delta G$) of the catalytic GeO oxidation [Eq. (3)], competing catalyst passivation [Eq. (4)], and potential "decomposition of the passivated catalyst" by the GeO through the following reaction:

$$SnO_2(s) + GeO(g) \rightarrow GeO_2(s) + SnO(g). \quad (5)$$

Based on the ease of occurrence (i.e., more negative $\Delta G$), the reactions are ranked as follows: reaction (4) $\gg$ reaction (3) $\gg$ reaction (5), indicating a thermodynamically preferred catalyst passivation over catalytic GeO oxidation, which conflicts with our assumption of a thermodynamically preferred GeO incorporation. The resulting positive Gibbs free energy of reaction (5), in comparison, confirms the stability of the passivated catalyst.

These partially conflicting results of thermochemistry and our growth experiment call for detailed *ab initio* studies to better understand the microscopic processes during the catalytic MBE growth of $GeO_2$ and likely other oxides.

### E. Control of $(Sn_xGe_{1-x})O_2$ alloy composition $x$

Irrespective of the underlying microscopic mechanisms, our experimental results shown in Fig. 7(a) demonstrate that increasing the oxygen flux significantly enhances Sn incorporation in $(Sn_xGe_{1-x})O_2$ alloys, a trend consistent with observations of increased In-incorporation in conventional-MBE-grown $(In_xGa_{1-x})O_2$.[35] In contrast to a decreasing In incorporation at increasing $T_G$ for the growth of $(In_xGa_{1-x})O_2$,[35] however, the Sn incorporation into $(Sn_xGe_{1-x})O_2$ increases in our case [see Fig. 4(b)]. We attribute the observed oxygen-flux-dependence to the catalyst passivation discussed above, whereas the increased SnO incorporation at increasing $T_G$ is likely related to a GeO desorption preempting its catalytic oxidation.

Notwithstanding, these observations suggest that both the oxygen flux and the $T_G$ are key parameters in determining the Sn and Ge composition in $(Sn_xGe_{1-x})O_2$, e.g., allowing the control of $x$ between 0.2 and 0.7 for a fixed SnO/(SnO + GeO) flux stoichiometry of $x_{flux} = 0.57$. Obtaining alloys with $x < 0.2$ or $x > 0.7$ should be possible by decreasing or increasing the flux stoichiometry, respectively, since we have shown the trivial binary cases $x_{flux} = 0$ and $x_{flux} = 1$ to result in the growth of $(Sn_xGe_{1-x})O_2$ layers with $x = 0$ and $x = 1$, respectively. Understanding these dependencies allows for precise control over the alloy composition, which is essential for tailoring its properties to suit specific applications.

## IV. CONCLUSION

In summary, we systematically explored the cation incorporation into $GeO_2$, $SnO_2$, and $(Sn_xGe_{1-x})O_2$ to enable controlled suboxide-MBE growth of these technologically important (ultra)wide-bandgap semiconductors. Cations that are not incorporated into the oxide film desorb as volatile suboxides GeO or SnO.

The following differences between conventional MBE (using cation fluxes) and S-MBE (providing suboxide fluxes instead) were observed for both $GeO_2$ and $SnO_2$:

- Conventional MBE enables oxide-layer etching (Ge can etch $GeO_2$; Sn or Ge can etch $SnO_2$) through suboxide formation

and desorption, which is not possible in S-MBE (e.g., SnO does not etch $SnO_2$ but just desorbs as SnO).

- In S-MBE, the growth-rate increases linearly with active oxygen flux, whereas growth only starts at a threshold oxygen flux that is required for intermediate suboxide formation on the growth front in conventional MBE.

Our comparative analysis of binary $GeO_2$ and $SnO_2$ growth kinetics revealed that Sn or SnO incorporation is kinetically favored over that of Ge or GeO in both conventional and S-MBE due to (i) the higher oxidation efficiency of Sn (30% compared to 6% for Ge) and SnO (23% compared to 3.3% for GeO) as well as (ii) the lower vapor pressure of SnO compared to that of GeO. As practical consequences, $SnO_2$ can be grown at higher substrate temperatures or lower active oxygen flux than $GeO_2$.

In the ternary $(Sn_xGe_{1-x})O_2$ system, the film composition $x$ can strongly deviate from the provided SnO/(SnO + GeO) flux stoichiometry. Despite the kinetic advantage for SnO incorporation in binary growth, GeO is favorably incorporated in the ternary growth. The observed enhancement of GeO incorporation coinciding with suppressed SnO incorporation in comparison to binary growth is tentatively explained by a catalytic effect synergizing the kinetic advantage of SnO oxidation with a thermodynamically or kinetically driven Ge–Sn cation exchange. Simple thermochemical calculations, however, are conflicting with a simple thermodynamically driven cation exchange, calling for future theoretical research to help understand the complex microscopics of the observed cation-exchange catalysis.

Nonetheless, higher $O_2$ fluxes and elevated growth temperatures promote SnO incorporation into $(Sn_xGe_{1-x})O_2$, and the flux stoichiometry can ultimately influence $x$ since the pure SnO and GeO fluxes were shown to result in the end-members of the alloy system, $SnO_2$ and $GeO_2$. These findings underscore the need for precise control over the growth environment beyond the mere suboxide flux stoichiometry to achieve the desired composition in $(Sn_xGe_{1-x})O_2$ thin films.

Our results experimentally and theoretically provide valuable insights for optimizing growth conditions to manage the competitive dynamics between GeO and SnO during $(Sn_xGe_{1-x})O_2$ ($0 \leq x \leq 1$) thin film growth by S-MBE. They are likely transferable to conventional MBE, where suboxide formation happens rapidly on the growth front,[20,36] as well as hybrid MBE, pulsed laser deposition, mist-, and metal-organic chemical vapor deposition.

## ACKNOWLEDGMENTS

We would like to thank Yongjin Cho for critically reading the manuscript; Emmanouil Kioupakis, Nadire Nayir, and Roger A. De Souza for the discussion; as well as Hans–Peter Schönherr and Claudia Hermann for the technical MBE support. This work was performed in the framework of GraFOx, a Leibniz-ScienceCampus partially funded by the Leibniz association. W.C. gratefully acknowledges the financial support from the Leibniz association under Grant No. K417/2021.

## AUTHOR DECLARATIONS
### Conflict of Interest

The authors have no conflicts to disclose.

### Author Contributions

**Wenshan Chen**: Data curation (equal); Formal analysis (equal); Investigation (equal); Methodology (equal); Writing – original draft (equal); Writing – review & editing (equal). **Kingsley Egbo**: Formal analysis (supporting); Investigation (supporting). **Joe Kler**: Formal analysis (supporting); Writing – review & editing (supporting). **Andreas Falkenstein**: Formal analysis (supporting); Writing – review & editing (supporting). **Jonas Lähnemann**: Formal analysis (supporting); Writing – review & editing (supporting). **Oliver Bierwagen**: Conceptualization (equal); Funding acquisition (equal); Project administration (equal); Supervision (equal); Writing – review & editing (equal).

## DATA AVAILABILITY

The data that support the findings of this study are available within the article and its supplementary material.